# A Bilateral River Bargaining Problem with Negative Externality

Shivshanker Singh Patel[1] • Parthasarathy Ramachandran[2]

**Abstract:** This article is addressing the problem of river sharing between two agents along a river in the presence of negative externalities. Where, each agent claims river water based on the hydrological characteristics of the territories. The claims can be characterized by some international framework (principles) of entitlement. These international principles are appears to be inequitable by the other agents in the presence of negative externalities. The negotiated treaties address sharing water along with the issue of negative externalities imposed by the upstream agent on the downstream agents. The market based bargaining mechanism is used for modeling and for characterization of agreement points.

## 1. Introduction

A large number of rivers flows across political boundaries with conflict arising out of the need to share the water flowing across the boundary in a fair and equitable manner. It has been estimated that there are some 267 water bodies (lakes and rivers) that spans across political borders, with Nile, Ganges, and Danube being some prominent examples (Giordano and Wolf, 2003). The increasing water scarcity makes the water in these rivers a source of conflict. Even rivers within a single national border is also contested by different cities, states, and use groups. Examples of such shared river courses that ended up as disputes include the Mekong, Indus and Nile. This problem is studied in the economics literature as the *river water sharing problem* (Ambec and Sprumont, 2002). A negotiated settlement has been widely advocated for resolving the conflict arising out of the river water sharing problem. Some examples of settlement arising out of established treaties or tribunals are:

  • Rio Grande: an interstate water sharing treaty between the states of Colorado, New Mexico, and Texas in the United States (Rio Grande compact)

[1] Indian Institute of Management Visakhapatnam, Andhra Pradesh, INDIA
   Email: shivshanker@iimv.ac.in
[2] Department of Management Studies
   Indian Institute of Science Bangalore, INDIA

• Cauvery river: national tribunal (by Government of India) to adjudicate the water sharing conflict between Karnataka and Tamilnadu state.

• Indus River: World Bank brokered treaty between India and Pakistan (Treaty for Common development of river basin)

Some other case studies about inter-national river treaties are also discussed by Barret (1994). Though these agreements and tribunal awards seek to resolve the disputes, they are not effective in removing the conflict completely. The conflict situations arise whenever there is a drought like situation leading to reduced flow and associated negative externalities. Hence, Kilgour and Dinar (1995) developed a water allocation model using the notion of transferable utility. This approach to the water sharing problem might overcome the often political nature of conflict as the agents might find it in their best interest to agree and implement the allocations. This notion of transferable utility and its use in the context of river water sharing problem was expanded later by others including Ambec and Sprumont (2002), Wang (2011), Dinar et al. (2010), and Brink et al.. (2012).

Pollution and floods are two major sources of negative externalities that originates from one agent but also impacts the other agents along the river stream caused by upstream agent to downstream agent in a river sharing scenario. In this article negative externality associated with pollution has been considered for analysis. We have underpinned the past literature in two aspects. Firstly, including the literature from the river sharing problem with respect to benefit maximization of agents. Later, we include the river sharing problem with induced negative externalities because of pollution etc. Next section provides brief literature review about the river sharing problem and associated negative externalities.

**2. Related Literature**

The claims of agents along a trans-boundary watercourses normally follows two principles,

• *Absolute Territorial Sovereignty (ATS) or The Harmon Doctrine:* It gives a riparian state full control over all waters generated within its territory, and can utilize those waters without considering dependent livelihood or claims of the other co-riparian agents. This principle is typically favored by the upper riparian agents.

• *Absolute Territorial Integrity (ATI):* This theory allows a downstream riparian state to demand the the full flow of the river from an upper riparian state without compromising the quantity or quality. Absolute territorial integrity logically make more sense to lower riparian agents.

These principles appeal to individual agents they contradict each other and hence may not be usable in practice. In general an agent's claim is always larger than their endowment. Agents' overlapping claims to river water make water a contested resource (Ansink and Weikard, 2009). The upstream agents who prefer the ATS disagree on the amount of water to be shared with downstream agents. Ambec and Sprumont (2002) proposed a compromise between the two doctrines by treating the ATS doctrine preferred by the upper riparian agent as a "core-like" constraint and the ATI doctrine favored by the lower riparian agent as "legitimate aspirations". The ATS and ATI principles addresses the right to use the river water without considering the responsibilities of the agents towards each other. Ni and Wang (2007) and Dong et al. (2012) later re-interpreted ATS and ATI to include the responsibilities of the agents towards each other.

The existing literature in the context of *river sharing problem* is influenced largely by research of Ambec and Sprumont (2002) and Ambec and Ehlers (2008). Their framework was described as a Transferable Utility (TU) game and considering a benefit function which is strictly concave with a single peak. Later, that motivated Wang (2011) to develop a bargaining framework to provide a mechanism for the transferable utility game of the river sharing problem. The bargaining framework discussed by Wang (2011) recognizes the existence of negative externalities but does not incorporate them in the developed framework.

The *River Sharing Problem* becomes more critical in the presence of negative externalities. In this context the negative externality is mostly in the form of pollution caused by an agent (upstream). Alcalde-Unzu et al. (2015) consider that when negative externalities of upstream agent is unknown, then the proposed clean-up cost vector gives an useful information for estimating limits in regard to the responsibility of each agent. Their results claim a cost allocation rule as Upstream Responsibility Rule (URR), which is claimed to be "fair". On the other hand, a non-cooperative water allocation between heterogeneous communities in an acyclic network of water sources is studied by Rebille and Richefort (2012). The solution is to impose a tax (optimal) on an agent that reflects the marginal damages and the marginal benefits that one agent transfer to the others.

Nevertheless, in case of trans-boundary rivers there are often no higher institutions that can enforce taxes on agents.

Therefore, it requires a cooperation and implementation of solution concepts that can provide optimal outcomes that is against free riding (Chander and Tulkens 1997). Ni and Wang (2007) takes the issue of river pollution and negative cost, and incorporated the principles of ATS and ATI in their analysis. They propose two methods to deal with the negative externalities due to pollution, namely the Local Responsibility sharing (LRS) method and Upstream Equal Sharing (UES) method. They provide an axiomatic characterization of these methods and claim that both the approach coincide with the Shapley value solution to the respective games.

Further, Dong et al. (2012) extended the work and proposed a new method of *Downstream Equal Sharing* (DES). These analysis are similar analytical explanation with respect to allocating negative cost as Ambec and Ehlers (2008) analyzed with respect to the water sharing and associated welfare allocation. Dong et al. (2012) also proposed a solution of *Shapley value* and the *core* within the framework of cooperative game theory. The VCG (Vickrey-Clarke-Groves) mechanism and Polluter Pay rules for allocating negative cost among the agents is also discussed in previous literature (Ambec and Ehlers, 2014).

In this paper, we have addressed the issue of negative externality that is imposed by an upstream agent on a downstream agents as pollution. The characterization of negative externality can be comprehended by the variables such as inflow of water, benefit associated with water usage, and a negative cost to mitigate the pollution. The Pigovian tax approach will not be easy to implement in the context of negative externality in a trans-boundary river sharing problem. This is due to the fact that there may not be a superior institutional or regulatory body that can enforce tax regime. Coase (1960) showed that when agents are affected by negative externalities efficient/equilibrium outcomes are possible through market mechanisms irrespective of their initial property rights allocations. In the case of two-agent River sharing problem the water inflow to an agent's territory defines their initial property rights according to the ATS doctrine. If we assume that this assignment of property right requires that they take the responsibility for the externalities caused by them, then the application of a bargaining framework between the two agents incorporating the negative externalities would result in an efficient outcome.

Hence, we propose that the market based negotiated treaty that accommodates negative externalities is appropriate. The upstream agent agrees to incorporate the negative cost in her benefit function to pollute the river and also gets opportunity of trading (selling) surplus water to downstream agent. On the other hand, downstream agent incorporates the cost of cleaning polluted water in her benefit function and also trading (buying) extra water from upstream. Both agents try to maximize their utility to reach Pareto optimality. The utility from consuming water incorporates negative externalities in order to account for the agents' behavior. We identify individually rational bargaining strategies for the two agents. Section 4 explains the characterization of negative externalities in the context of river sharing. Section 5 develops the 2-agent river bargaining problem with induced negative externalities followed by solutions discussion in Section 6-8.

## 4. Negative Externality

For a given pollution level the cost of extracting water of a certain stated quality for the lower riparian state decreases with the increase discharge from the upper riparian state. But, the increase in discharge could result in flood damages beyond a certain level. From the perspective of the lower riparian state the total cost need to be minimized. In figure 1 the total cost curve due to negative externalities is represented. It is a convex curve with the trough being the region preferred by the lower riparian states. As the pollution level increases it can be surmised that the total cost curve will move upwards (dashed curves).

Even in the absence of any external pollution added to the stream by the upper riparian state, the lower riparian state would incur a certain cost for extracting water due the fact of decreasing flow to the downstream will increase the pollution density. In Fig. 1 this is represented by the solid line. $C^*$ represent the absolute minimum cost necessary to extract water from the river by the lower riparian state. Any increase in cost beyond this level can be attributed as the negative externality imposed by the upper riparian agent on the lower riparian agent.

Applying the rights with responsibility principle, the lower riparian agent would expect to be compensated for this negative externality. The following sections introduces these negative externalities in the bargaining framework for the river sharing problem.

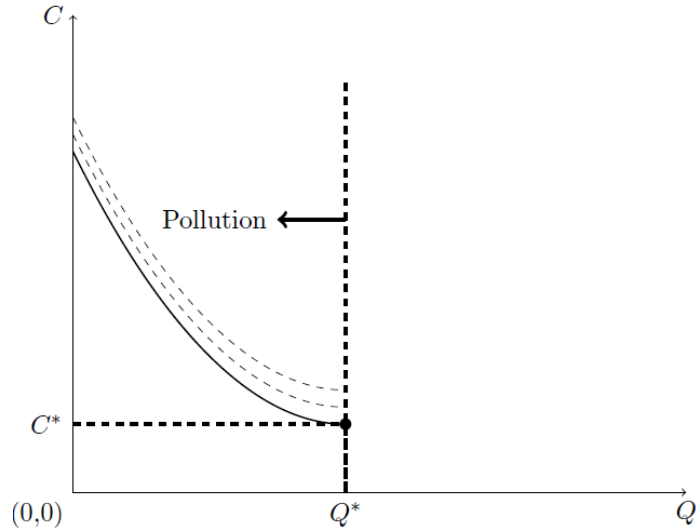

Fig. 1: Negative externalities put on downstream agent in river sharing problem

## 5. Bargaining Framework and Modeling

In this section, we have used a general model for the river sharing problem between two agents (see Fig.2) with negative externalities. Consider a pair of agents {1,2} where 1 is upstream from 2 as depicted in Fig. 2. Let $e_i, i \in \{1,2\}$ be the endowment of agent $i$ to the river based on the water originating within their territory. It is assumed that $e_1$ and $e_2$ are spatially independent of each other. Let $x_1$ and $x_2$ be the amount of water consumed by the two agents.

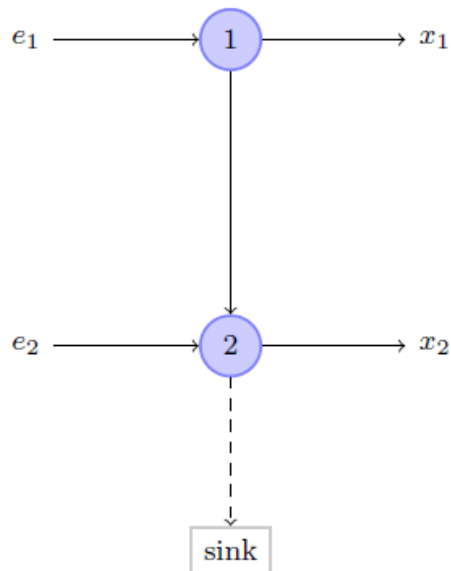

Fig. 2: River sharing problem for 2-agents (arrow shows the direction of water flow

In the general river sharing problem discussed by Ambec and Ehlers (2008) the case of satiable agents where marginal benefit of downstream agent is higher than the upstream agent is studied. Without loss of generality, both agents (1 and 2) try to maximize their individual benefit. In order to do that either they will follow ATS or ATI. ATS is the core lower bound for the agents Ambec and Ehlers (2008). However, they can maximize the total benefit by transferring the water from lower marginal benefit to higher marginal benefit agent (upstream to downstream). And, with an appropriate mechanism the downstream agent can transfer utility to upstream agent against the traded water by the upstream agent. This approach will motivate upstream agent not to follow the ATS and look for some alternative negotiated treaty by which an upstream agent can trade water with the downstream agent in order to maximize her individual benefit. In such a case, for a negotiated treaty and water trading it is necessary to have a market for cooperation where agents can trade to maximize their total and individual utilities with a bargaining mechanism for the two agents to trade (Wang, 2011).

When there is water traded between the agents, let $t_1$ be the money transferred by agent 2 to agent 1 in exchange for the transferred water quantity $e_1 - x_1$. The utility functions for both the agents by consuming water is as given below.

$$u_1(x_1, t_1) = b_1(x_1) + t_1 \qquad (1)$$

$$u_2(x_2, t_2) = b_2(x_2) + t_2 \qquad (2)$$

In the above given equations, the functions $b_1, b_2: \mathfrak{R}_+ \to \mathfrak{R}$ are the benefit functions for agent 1 and 2. It is assumed to be strictly concave and differentiable at every point $x_1, x_2 > 0$ and further $x_1 \leq e_1, x_2 \leq e_1 - x_1 + e_2$. If agent 1 receives any payment $t_1$ then for agent 2, then for agent 2, $t_2 = -t_1$.

A 2 − *agent river sharing problem* with trading can be represented by tuple $<2, e, b>$, where $e = (e_1, e_2)$ and $b = (b_1, b_2)$. The allocation vector $(x_1, x_2, t_1) \in \mathfrak{R}^3$ specifies the amount of water allocated to the two agents and the money transferred from agent 2 to agen 1 such that $x_1 + x_2 \leq e_1 + e_2$. Also, $t_2 = -t_1$ implies that there is no transaction cost in this model. Then [Wang, 2011] provide the following equilibrium condition

$$b_{1,}(x_1) = b_{2,}(x_2) \qquad (3)$$

for bilateral trading to happen. This model does not incorporate the impact of the negative externalities discussed earlier.

Our aim is to incorporate the notion of negative externalities in the river sharing problem in a bargaining model. The negative externality due to pollution caused by any upstream agent to the downstream agent is considered here. The downstream agent faces an increased cost for extracting some stated quality of water. As we have discussed earlier ATS principle is usually preferred by the upstream agent with the belief that it will maximize her utility. This claimed right and the responsibility for not imposing any negative externality on the downstream agent, it can induce an environment for trading to maximize the individual and total utility.

By assigning initial property rights using the ATS doctrine and invoking the *Coase theorem* it is known that a bargaining model (market mechanism) would lead to efficient outcomes, provided that there are no private pieces of information (Patrick, 2001). The information elements of the river sharing problem are the individual endowments $e_1$ and $e_2$ and the pollution caused by the economic activities of the upstream agent. The endowments $e_1$, and $e_2$ are assumed to be the initial property rights of agents 1 and 2 respectively. It can be assumed to be public information as monitoring stations can be established and jointly operated or by other neutral bodies. Similarly the pollution levels can be monitored and measured using established standards. Hence, the $2-$ *agent river sharing problem with negative externalities* has clear initial property rights and with perfect information on endowments and pollution levels.

In order to encourage agents to participate in a negotiation and trading the individual utilities of the agents should be more than what they can achieve individually. There should be *individual rationality* for the agents. That will motivate any agent to participate in trading and bargaining.

It is necessary to incorporate a cost component to upstream agent for her responsibility for imposing negative externalities on downstream agent. This cost (penalty) possibly formalized through treaty agreements could take the form of additional discharge to the downstream agent (viewed as negative water for the upstream agent). This approach introduces incentive and threat to both the agents. The incentive part is the transferable utility for trading additional discharge from downstream agent to the upstream agent. And, threat as negative externalities in the form of compensatory additional discharge from the upstream agent to the downstream agent.

Aforementioned framework of addressing negative externality in 2- *agent river sharing problem* can be expressed by following mathematics structure.

The agents $i \in \{1,2\}$ have benefit functions $b_i: \Re_+ \to \Re$ such that $b_i(0) = 0$ and it is differentiable for all $x_i > 0$. Also there exists a satiation point $x_i^*$ such that $b_i(x_i^*) > b_i(x_i)$, for all $x_i \neq x_i^*$, $b'_i(x_i^*) = 0$ and $b''_i(x_i^*) < 0$.

For agent 2, let $c_2(E_2, x_2): \Re_+^2 \to \Re_+$ be the cost of accessing water quantity $x_2$ when $E_2$ is the water available for consumption. The negative externality introduced by the first agent can be measured in water equivalent terms (negative water) that can be the equivalent of the cost to be incurred by the second agent for extracting water of a certain stated quality. This cost imposed by the first agent on the second agent will be a function of the quantity allocated and consumed by the first agent and her total endowment. We make the following assumption about the nature of this cost.

For given level of pollution added by agent 1 to the stream, the negative externality cost imposed on the second agent $c_1^w(e_1, x_1): \Re_+^2 \to \Re_+$ and $0 \leq c_1^w < e_1$. The nature of this function is such that for a given level of endowment $e_1 = e$, $c_1^{w\prime}(e, x_1) = c_{1w}$ and $c_1^w(e, 0) = 0$.

This assumption introduces a direct negative cost (penalty) in terms of "negative water" for the negative externality caused for every unit of freshwater consumed by agent 1. This negative water is a penalty which cuts her upper bound for consuming maximum amount of water.

In previous mechanisms given in literature, an agent 1 could consume maximum upto $e_1$ but with new mechanism she can only consume $e_1 - c_1^w(e_1)$. However, if agent 1 consumes zero amount of water in such situation she can sell whole of $e_1$ to the downstream agent. But, if she consumes $e_1 - c_1^w(e_1)$ she cannot trade any water. The penalty of $c_1^w(e_1)$ is the compensation to the downstream agent due to the negative externality imposed on her.

Next, in section 6 we define the 2-person bargaining problem that explains incorporating the notion of negative water in to utility function of the two agents.

## 6. Formulation

The river sharing problem with negative externalities can be modeled with transferable utility (TU) market based mechanism as bargaining problem. The bargaining model incorporating the notions of transferable utility, negative cost and also the associated opportunity cost for transferring a negative externality from one agent to another is explained below.

The two agent river bargaining problem is represented by the 4-tuple $< 2, e, c_w, \alpha >$. In a 2 agent bilateral river bargaining problem, $e = \{e_1, e_2\}$ are the water endowments for the agents within her territory, $c_w = \{c_{1w}\}$ is the negative water penalty on upstream agent (1) for generating negative externalities for the downstream agent, and $\alpha$ is the transferable side payment from agent 2 to agent 1 for every unit water traded.

The river sharing problem has widely accepted the assumption (2) that the benefit function is strictly concave. The Eqs. 4 and 5 shows the utility function[3] of agents 1 and 2 respectively that gives rationality to the agents to bargain or not in evaluating the allocation vector **x** and the transferable utility $\alpha$.

If agent 1 is consuming $x_1$ units of water, then her utility is given by

$$z_1(x_1) = \overbrace{a_1 x_1 - \frac{b_1}{2} x_1^2}^{benefit} + \overbrace{\alpha_1(e_1 - x_1(1 + c_{1w}))}^{TU} + \overbrace{\beta_1 x_1 c_{1w}}^{opportunity\ cost} \quad . \tag{4}$$

In the Eq. 4, $c_{1w}$ is the penalty negative water with respect to negative externality caused by agent 1 on agent 2. The $(e_1 - x_1(1 + c_{1w}))$ is *surplus water* or conserved water and that can be traded by agent 1 to agent 2 for generating extra revenue. If $\alpha_1$ is the value associated by agent 1 per unit of water then the revenue expected from the trade is $\alpha_1(e_1 - x_1(1 + c_{1w}))$. The other benefit enjoyed by agent 1 is the cost of not having to bring the water quality to acceptable levels by the downstream agent. This is an opportunity cost not incurred by agent 1 and hence is an additional benefit to her and is given by $\beta_1 x_1 c_{1w}$. If agent 1 follows ATS then she has to bear the cost of pollution. However, in a bilateral trading of water she can release some amount of polluted water and enjoy the savings from associated opportunity cost. Similarly, the utility function for agent 2 represented as,

---

[3] As we have mentioned earlier through this river bargaining problem we would like to incorporate the negotiated treaty of sustainable nature that capture the negative externalities of upstream agent and motivates both the agents to participate in the trade to be better off. It has been noticed that in two person bargaining problem sometimes one player's payoff increases as the disagreement payoff to other player decreases. So it gives an incentive for a player to have more favorable disagreement payoffs. In a river bargaining problem with negative externalities it appears that the upstream agent does have more favorable disagreement points due to the penalty of negative water against her generated negative cost to downstream flow. This notions are taken in to account while writing the utility function of the agent

$$z_2(x_2) = a_2 x_2 - \frac{b_2}{2} x_2^2 - \alpha_2(x_2 - e_2 - x_1 c_{1w}) - \beta_2 x_1 c_{1w}. \tag{5}$$

In Eq. 5, the term $\alpha_2(x_2 - e_2 - x_1 c_{1w})$ represents a transferable utility and term $\beta_2 x_1 c_{1w}$ is the cost incurred by agent 2 in extracting quality water. The *transferable utility* and *opportunity cost* have negative signs because these are actual expenses incurred by agent 2. Being rational agents the agents will seek to maximize their utilities giving raise to a decision problem. The agents maximize their utility by choosing appropriate consumption levels $x_1$ and $x_2$ and agreeing on the transferable utility $\alpha$. The analysis of this bargaining between agents over $\alpha = \alpha_1 = \alpha_2$ is discussed in following section (§ 7).

## 7. Solution

The two agent river bargaining problem is a form of game with transferable utility. This two person bargaining problem characterized by three parameters: disagreement payoffs of agent 1, disagreement payoff of agent 2 and the transferable utility on which the agents will agree upon to trade [Myerson, 1991].

Over the structure of the two-person river bargaining problem both the agents will be faced with the following decision problem (Eq. 6 and Eq. 7 respectively for agents 1 and 2).

$$\max z_1 \tag{6}$$

$$\text{s.t. } x_1 \leq e_1$$

at the same time agent 2 solves Eq. 7

$$\max z_2 \tag{7}$$

$$\text{s.t. } x_2 \geq e_2$$

### 7.1 Characterization

*Axiom1 [Individual Rationality].* An agent will take part in a bargaining for trading any economic commodity if only if when she is better off in participation.

*Proposition 1.* In the bilateral river sharing problem $< 2, e, c^w, \alpha >$ with negative externalities if the utility function of the upstream agent is expressed by Eq. 4, the upstream agent will participate in the trade only if $\frac{a_1 + \beta c_{1w}}{b_1} \geq \frac{e_1}{1 + c_{1w}}$.

*Proof.* By *Axiom1* of individual rationality, agent 1 will try to maximize her utility by maximizing the utility function given in 4. The first order necessary condition for maximization of Eq. 4 is as follows.

$$\frac{\partial z_1}{\partial x_1} = a_1 - b_1 x_1 - \alpha_1(1 + c_{1w}) + \beta_1 c_{1w} = 0$$

$$x_1 = \frac{a_1 - \alpha(1 + c_{1w}) + \beta_1 c_{1w}}{b_1}$$

By assumption 2 the above stationary point will maximize her utility function. The constraint in the decision problem (6) gives raise to the following (eq. 8) optimal consumption level of the first agent.

$$x_1^* = \min\left\{\frac{a_1 - \alpha_1(1 + c_{1w}) + \beta_1 c_{1w}}{b_1}, \frac{e_1}{1 + c_{1w}}\right\}$$

(8)

The agent 1 will only participate in any trading and not follow ATS only when (due to axiom 7.1),

$$z_1(x_1) \geq z_1(e_1/(1 + c_{1w})) \qquad (9)$$

Substituting the consumption levels in the utility function (eq. 4)

$$a_1 x_1 - \frac{b_1}{2} x_1^2 + \alpha_1(e_1 - x_1(1 + c_{1w})) + \beta_1 x_1 c_{1w} \geq a_1\left(\frac{e_1}{1+c_{1w}}\right) - \frac{b_1}{2}(e_1/(1+c_{1w}))^2 + \beta_1(e_1/(1+c_{1w}))c_{1w}$$

(10)

Simplifying the expression yields

$$a_1\left[x_1 - \frac{e_1}{1 + c_1 w}\right] - \frac{b_1}{2}\left[x_1^2 - \left(\frac{e_1}{1 + c_{1w}}\right)^2\right] + \alpha_1(e_1 - x_1(1 + c_{1w})) + \beta_1 c_{1w}\left[x_1 - \frac{e_1}{1 + c_{1w}}\right]$$
$$\geq 0$$

$$a_1 - \frac{b_1}{2}\left[x_1 + \frac{e_1}{1+c_{1w}}\right] - \alpha_1(1+c_{1w}) + \beta_1 c_{1w} \geq 0$$

$$\frac{a_1 - \alpha_1(1+c_{1w}) + \beta_1 c_{1w}}{b_1} \geq \frac{1}{2}\left[x_1 + \frac{e_1}{1+c_{1w}}\right]$$

substitution of $x_1^* = \frac{a_1 - \alpha_1(1+c_{1w}) + \beta_1 c_{1w}}{b_1}$ from Eq.8 gives

$$\frac{a_1 - \alpha_1(1 + c_{1w} + \beta_1 c_{1w})}{b_1} \leq \frac{e_1}{1+c_{1w}} \tag{11}$$

Hence,

$$\alpha_1 \geq -\frac{b_1 e_1}{(1+c_{1w})^2} + \frac{\beta_1 c_{1w}}{1+c_{1w}} + \frac{a_1}{1+c_{1w}} \tag{12}$$

This is the lower limit ($\alpha_1^l$) of the TU that is acceptable to agent 1. Since $\alpha_1 \geq 0$

$$\frac{a_1 + \beta c_{1w}}{b_1} \geq \frac{e_1}{1+c_{1w}} \tag{13}$$

*Proposition 2.* In the bilateral river sharing problem $< 2, e, c^w, \alpha >$ with negative externalities if the utility function of the downstream agent is expressed by Eq. 5, the downstream agent will participate in trade only when $\frac{a_2 + \beta_2 c_{1w}}{b_2} \geq e_2 + \frac{e_1 c_{1w}}{1+c_{1w}}$.

*Proof.* The agent 2 will also follow her utility function (Eq. 7) and try to maximize it. Given the endowments, if agent 2 is allocated $x_2$, agent 1 will be allocated $x_1 = e_1 + e_2 - x_2$. Then the first order necessary condition for optimality gives raise to,

$$\frac{\partial z_2}{\partial x_2} = a_2 - b_2 x_2 - \alpha_2(1+c_{1w}) + \beta_2 c_{1w} = 0$$

$$x_2 = \frac{a_2 - \alpha_2(1+c_{1w}) + \beta_2 c_{1w}}{b_2} \tag{14}$$

The constraint on the decision problem of the second agent gives raise to

$$x_2^* = \max\left\{e_2 + c_{1w} x_1, \frac{a_2 - \alpha_2(1+c_{1w}) + \beta_2 c_{1w}}{b_2}\right\} \tag{15}$$

By *Axiom1 of individual rationality,* the agent 2 will also participate in any trading only if

$$z_2(x_2) \geq z_2(e_2 + c_{1w}x_1) \tag{16}$$

Substituting the consumption levels in the utility function Eq. 5

$$a_2 x_2 - \frac{b_2}{2} x_2^2 - \alpha_2(x_2 - e_2 - x_1 c_{1w}) - \beta_2 c_{1w}(x_1)$$

$$\geq a_2(e_2 + c_{1w}x_1) - \frac{b_2}{2}(e_2 + c_{1w}x_1)^2 - \beta_2 c_{1w}(x_1)$$

$$a_2(x_2 - (e_2 + c_{1w}x_1)) - \frac{b_2}{2}(x_2^2 - (e_2 + c_{1w}x_1)^2) - \alpha_2(x_2 - (e_2 + x_1 c_{1w})) \geq 0.$$

Simplifying yields

$$\frac{a_2 - \alpha_2}{b_2} \geq \frac{1}{2}(x_2 + e_2 + c_{1w}x_1).$$

Substitution of $x_2^* = \frac{a_2 - \alpha_2(1 + c_{1w}) + \beta_2 c_{1w}}{b_2}$ from Eq. 15 gives

$$\frac{a_2 - \alpha_2(1 + c_{1w}) + \beta_2 c_{1w}}{b_2} \geq e_2 + c_{1w}x_1$$

Substituting $x_1 = e_1 + e_2 - x_2$ in the above equation and replacing $x_2$ by $x_2^* = \frac{a_2 - \alpha_2(1 + c_{1w}) + \beta_2 c_{1w}}{b_2}$ yields

$$\frac{a_2 - \alpha_2(1 + c_{1w}) + \beta_2 c_{1w}}{b_2} \geq \frac{e_2(1 + c_{1w}) + e_1 c_{1w}}{(1 + c_{1w})}$$

$$\alpha_2 \leq \frac{a_2 + \beta_2 c_{1w}}{1 + c_{1w}} - \frac{b_2}{1 + c_{1w}}\left(e_2 + \frac{e_1 c_{1w}}{1 + c_{1w}}\right) \tag{17}$$

This is the upper limit ($\alpha_2^u$) of the TU that is acceptable to agent 2. Since $\alpha_2 \geq 0$, hence

$$\frac{a_2 + \beta_2 c_{1w}}{b_2} \geq e_2 + \frac{e_1 c_{1w}}{1 + c_{1w}} \tag{18}$$

### 7.2 Feasibility of agreement

*Lemma 3.* In the bilateral river sharing problem $< 2, e, c^w, \alpha >$ with negative externalities with the agent utilities expressed by Eqs. 4, 5 the two agents will have an agreement point only if

$$\frac{a_1+a_2}{1+c_{1w}} + c_{1w}(\beta_1 + \beta_2) \geq \frac{e_2+e_1}{1+c_{1w}} / (\frac{1}{b_1} + \frac{1}{b_2}) \tag{19}$$

*Proof.* The utility transferred by agent 2 to agent 1 is a sufficient condition for trading must appear when

$$\alpha_2(x_2 - e_2 - c_{1w}x_1) = \alpha_1(e_1 - x_1 - c_{1w}x_1)$$

For agreement between the agents $\alpha_1 = \alpha_2$, and hence

$$x_2 - e_2 - \frac{e_1 c_{1w}}{1+c_{1w}} = \frac{e_1}{1+c_{1w}} - x_1.$$

Substituting $x_2^*$ and $x_1^*$ and $\alpha_1 = \alpha_2$ yields

$$\frac{a_2 - \alpha(1 + c_{1w} + \beta_2 c_{1w})}{b_2} - e_2 - \frac{e_1 c_{1w}}{1+c_{1w}} = \frac{e_1}{1+c_{1w}} - \frac{a_1 - \alpha(1 + c_{1w} + \beta_1 c_{1w})}{b_1}$$

Simplifying the above expression yields

$$\alpha^* = \frac{a_1 + a_2 - (e_1+e_2)/(\frac{1}{b_1}+\frac{1}{b_2}) + c_{1w}(\beta_1+\beta_2)}{1+c_{1w}}$$

(20)

since $\alpha \geq 0$, hence

$$\frac{a_1+a_2}{1+c_{1w}} + c_{1w}(\beta_1 + \beta_2) \geq \left(\frac{e_2+e_1}{1+c_{1w}}\right) / \left(\frac{1}{b_1} + \frac{1}{b_2}\right) \tag{21}$$

### 7.3. Sufficiency for agreement

*Corollary 1.* In the bilateral river sharing problem $< 2, e, c^w, \alpha >$ with negative externalities the agents will arrive at an agreement iff $\alpha_1^l \leq \alpha^* \leq \alpha_2^u$.

*Proof.* The proof is rather direct by the application of Propositions 1, 2 and Lemma 1.

### 8. Numerical Illustration

The bargaining model is numerically illustrated by assuming the following parameters: $a_1 = 4$ $b_1 = .02$, $a_2 = 2$, $b_2 = .04$, $\beta_1 = .02$, $\beta_2 = .2$, $c^w = 4$ $e_1 = \delta e_2$. Where $\delta \in \{0,1,2,\ldots 30\}$. The lower ($\alpha_1^l$) and upper ($\alpha_2^u$) bounds of the TU and the agreement point ($\alpha^*$) derived earlier are plotted in Fig. 3. The specific case of $\delta = 30$ is represented in Fig. 4. As the figures show, there exists a specific specific region of the endowment in which the agents are able to agree on the TU value and engage in trade. Disagreement Lower values of the endowment

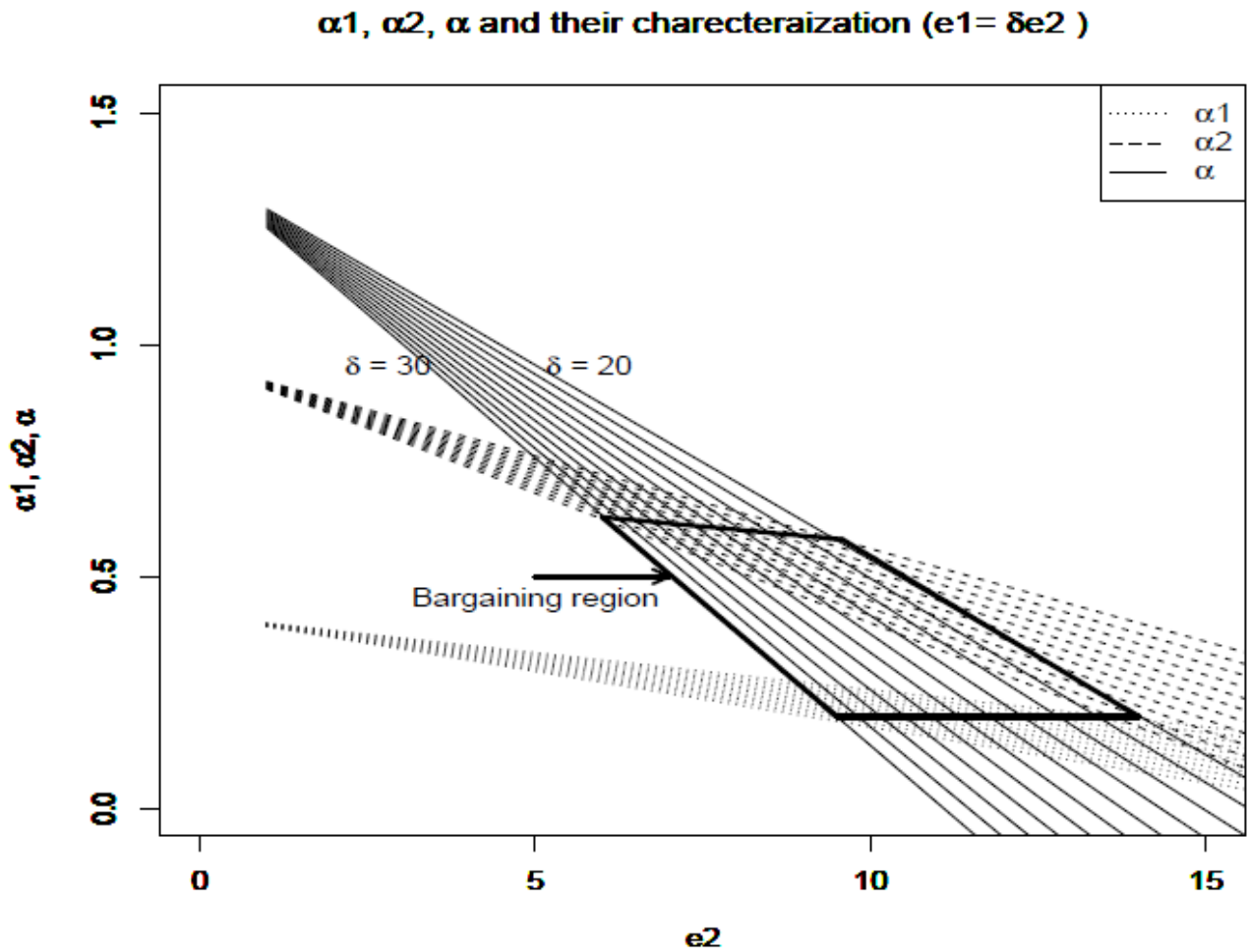

Fig. 3: Representation of agreement points and disagreement points

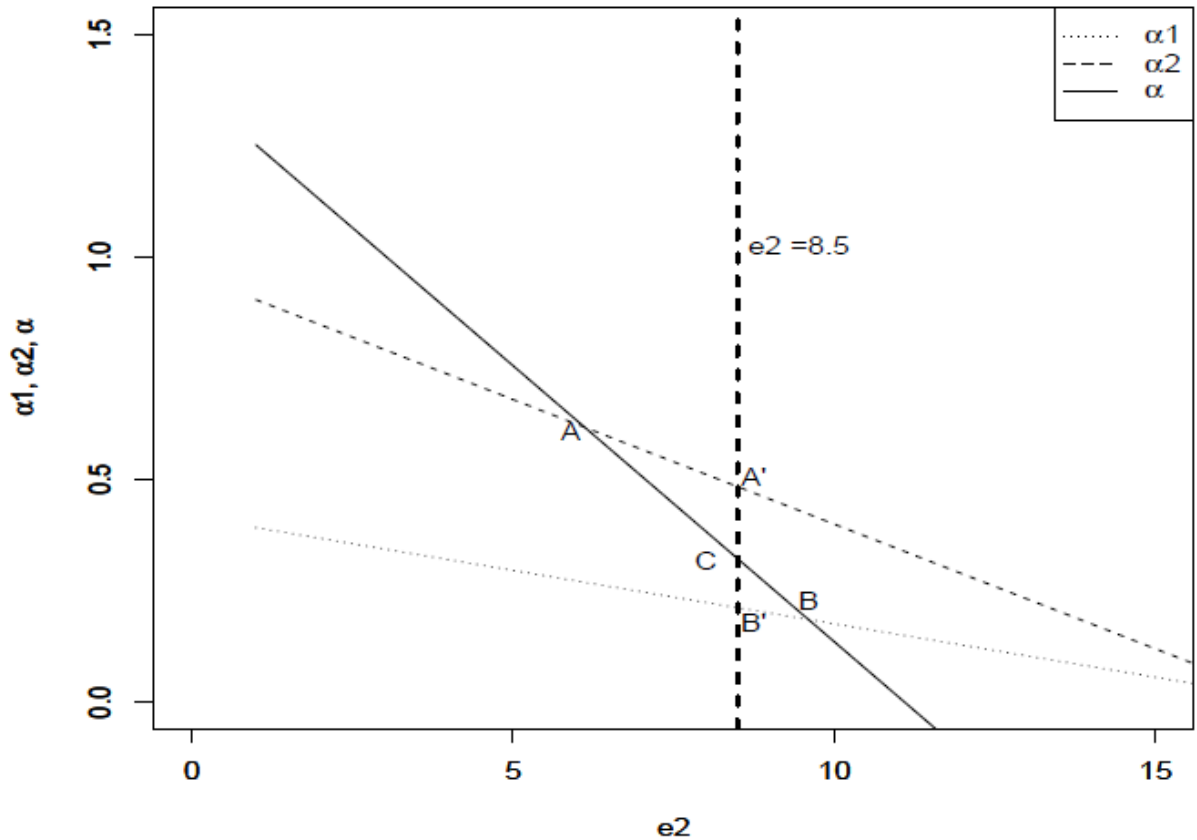

Fig. 4: Representation of agreement points and disagreement points for $\delta = 30$

Fig. 3 shows the feasible region for bargaining between two agents for agreements and Fig. 4 represents the $\delta = 30$. If we see Fig. 3, it is found that for $\delta = 20$ to $\delta = 30$ there is solution is existing which translates a bargaining solution for two agents. It can be also understood that for a given $e$ and $\delta$ there is a unique bargaining solution. The Fig. 4 is more enlarged view of Fig. 3 for $\delta = 30$.

In Fig. 4, $A$ represents the $(\alpha_2^u)$ and $B$ depicts the value of $(\alpha_1^l)$ for a feasible along the increasing value of $e_2$ along the $x - axis$. The $(\alpha^*)$ is represented by $C$ and it can be obtained a straight line passing through $e_2 = k$ where, $k$ is any value for which $(\alpha^*)$ lie between the $\alpha_2(A')$ and $\alpha_1(B')$.